\magnification=1200
\baselineskip=15pt

\def\P{\Phi}
\def\L{\Lambda}
\def\a{\alpha}
\def\b{\beta}

\def\d{\delta}
\def\l{\lambda}
\def\N{{\cal N}}
\def\G{{\cal G}}
\def\F{{\cal F}}
\def\half{{1\over 2}}
\def\tr{{\rm tr}}

\rightline{UCLA/99/TEP/8}
\rightline{Columbia/99/Math}

\bigskip

\centerline{{\bf SEIBERG-WITTEN THEORY AND INTEGRABLE SYSTEMS}
\footnote*{Research supported in part by the 
National
Science Foundation under grants PHY-95-31023 and DMS-98-00783.}}

\bigskip
\bigskip

\centerline{\bf Eric D'Hoker${}^1$ and D.H. Phong${}^2$}

\bigskip

\centerline{${}^1$ Department of Physics}
\centerline{University of California, Los Angeles, CA 90095}

\medskip

\centerline{${}^2$ Department of Mathematics}
\centerline{Columbia University, New York, NY 10027}

\bigskip

\centerline{\bf Abstract}

\bigskip

We summarize recent results on the resolution of two intimately related 
problems, one physical, the other mathematical. The first deals with the 
resolution of the non-perturbative low energy dynamics of certain $\N=2$ 
supersymmetric Yang-Mills theories. We concentrate on the theories with one 
massive hypermultiplet in the adjoint representation of an arbitrary gauge 
algebra $\G$. The second deals with the construction of Lax pairs 
with spectral parameter for certain classical mechanics ``Calogero-Moser" 
integrable systems associated with an arbitrary Lie algebra $\G$. We review the 
solution to both of these problems as well as their interrelation.

\bigskip
\bigskip

\centerline{{\it Based on Lectures delivered at }}
\centerline{{\it ``Integrability, the Seiberg-Witten and Whitham Equations",
Edinburgh, September 1998,}}
\centerline{{\it ``Workshop on Gauge Theory and Integrable Models", Kyoto,
January 1999,}} 
\centerline{{\it ``Supersymmetry and Unified Theory of Elementary Particles",
Kyoto, February 1999. }}

\vfill\break

\centerline{\bf I. INTRODUCTION}

\bigskip

Some of the most important physical problems of 
contemporary theoretical physics concern the behavior 
of gauge theories and string theory at strong coupling. 
For gauge theories,
these include the problems of confinement of color,
of dynamical chiral symmetry breaking, of the strong
coupling behavior of chiral gauge theories, of the 
dynamical breaking of supersymmetry. In each of these areas, 
major advances have 
been achieved over the past few years, and a useful 
resolution of some of these difficult problems
appears to be within sight.
For string theory, these include the problems of 
dynamical compactification of the 10-dimensional
theory to string vacua with 4 dimensions and of 
supersymmetry breaking at low energies. Already, it has 
become clear that, at 
strong coupling, the string spectrum is radically 
altered and effectively derives from the unique 
11-dimensional M-theory.

\medskip

This rapid progress was driven in large part by
the Seiberg-Witten solution of $\N=2$ supersymmetric 
Yang-Mills theory for $SU(2)$ gauge group [1] and by
the discovery of $D$-branes in string theory. 
Some of the key ingredients 
underlying these developments are 

\medskip

\item{(1)} Restriction to solving for the low energy
  behavior of the non-perturbative dynamics,
  summarized by the low energy effective action of
  the theory. 
    
\item{(2)} High degrees of supersymmetry. This  
has the effect of imposing certain holomorphicity constraints 
on parts of the low energy effective action, and thus of 
restricting its form considerably. For gauge theories in 4-dimensions,
we distinguish the following degrees of supersymmetry.

\itemitem{$\bullet$} $\N=1$ supersymmetry supports chiral fermions 
   and is the starting point for the Minimal Supersymmetric 
   Standard Model, the simplest extension of the Standard
   Model to include supersymmetric partners.

\itemitem{$\bullet$} $\N=2$ supersymmetry only supports non-chiral
   fermions and is thus less realistic as a particle physics 
   model, but appears better ``solvable". This is where the 
   Seiberg-Witten solution was constructed.

\itemitem{$\bullet$} $\N=4$ is the maximal amount of
  supersymmetry, and a special case of $\N=2$ supersymmetry
  with only non-chiral fermions and vanishing renormalization
  group $\beta$-function. Dynamically, the latter theory is the
  simplest amongst 4-dimensional gauge theories, and offers
  the best hopes for admitting an exact solution.

\item{(3)} Electric-magnetic and Montonen-Olive duality. 
  The free Maxwell equations are invariant under electric-
  magnetic duality when $\vec{E} \to \vec{B}$ and $\vec{B} \to 
  -\vec{E}$. In the presence of matter,  duality will require the presence of
both
  electric charge $e$ and magnetic monopole charge $g$ whose
  magnitude is related by Dirac quantization
   $e\cdot g \sim \hbar$. Thus, weak
   electric coupling is related to large magnetic coupling.
   Conversely, problems of large
   electric coupling (such as confinement of the color electric 
   charge of quarks) are mapped by duality into problems
   of weak magnetic charge. It was conjectured by Montonen
   and Olive that the $\N=4$ supersymmetric Yang-Mills theory
   for any gauge algebra $\G$ is mapped under the interchange 
   of electric and magnetic charges, i.e. under 
   $e \leftrightarrow 1/e$ into the theory with dual gauge algebra $\G^\vee$.
When combined with the
   shift-invariance of the instanton angle $\theta$ this
   symmetry is augmented to the duality group $SL(2,{\bf Z})$,
   or a subgroup thereof.
   
\item{(4)} Finally, the Maldacena equivalence  
   between Type IIB superstring theory on $AdS_5 \times S^5$
   and 4-dimensional $\N=4$ superconformal Yang-Mills theory
   is conjectured to hold at strong coupling. The AdS/SCFT
   correspondence thus establishes a link between certain
   non-perturbative phenomena in string theory and in gauge
   theory. 
   
\medskip

Of central interest to many of these exciting developments
is the 4-dimensional supersymmetric Yang-Mills theory
with maximal supersymmetry, $\N=4$, and with arbitrary gauge 
algebra $\G$. Here, we shall consider a
generalization of this theory, in which a mass
term is added for part of the $\N=4$ gauge multiplet,
softly breaking the $\N=4$ symmetry to $\N=2$.  
As an $\N=2$ supersymmetric theory, the theory has a $\G$-gauge
multiplet, and a hypermultiplet in the adjoint representation 
of $\G$ with mass $m$. 

\medskip

This generalized theory enjoys many of the same properties
as the $\N=4$ theory : it has the same field contents;
it is ultra-violet finite;
it has vanishing renormalization group $\beta$-function,
and it is expected to have Montonen-Olive duality symmetry.
For vanishing hypermultiplet mass $m=0$, the $\N=4$ theory
is recovered. For $m\to \infty$, it is possible to choose
dependences of the gauge coupling and of the gauge scalar
expectation values so that the limiting theory is one of many
interesting $\N=2$ supersymmetric Yang-Mills theories.
Amongst these possibilities for $\G=SU(N)$ for example, 
are the theories with any number of hypermultiplets in the
fundamental representation of $SU(N)$, or with product
gauge algebras $SU(N_1)\times SU(N_2) \times \cdots \times SU(N_p)$,
and hypermultiplets in fundamental and 
bi-fundamental representations of these product algebras.
   
\medskip

Remarkably, the Seiberg-Witten theory for $\N=2$ supersymmetric
Yang-Mills theory for arbitrary gauge algebra $\G$ appears 
to be intimately related with the existence of certain  
classical mechanics integrable systems. This relation was
first suspected on the basis of the similarity between the Seiberg-Witten
curves and the spectral curves of certain integrable models [2].
Then, arguments were developed that Seiberg-Witten theory 
naturally produces integrable structures [3]. But a
connection derived from first principles
between Seiberg-Witten theory and 
integrable models still seems to be lacking.

\medskip

For the $\N=2$ supersymmetric Yang-Mills theory with massive
hypermultiplet, the relevant integrable system appears to be
the {\it elliptic Calogero-Moser system}. For $SU(N)$ gauge group, 
Donagi and Witten [3] proposed that the spectral curves of the 
$SU(N)$ Hitchin system should play the role of the Seiberg-Witten 
curves. Krichever (in unpublished work), Gorsky and Nekrasov,
and Martinec [4]
recognized that the $SU(N)$ Hitchin system spectral curves
are identical to those of the $SU(N)$ elliptic Calogero-Moser
integrable system. That the $SU(N)$ elliptic Calogero-Moser
curves (and associated Seiberg-Witten differential) do
indeed  provide the Seiberg-Witten solution for the $\N=2$
theory with one massive hypermultiplet was fully established 
by the authors in [5], where it was shown that

\medskip

\item{(1)} the resulting effective prepotential $\F$ (and thus the 
low energy effective action) reproduces correctly the logarithmic
singularities predicted by perturbation theory;

\item{(2)} $\F$ satisfies a renormalization group type equation which
determines explicitly and efficiently instanton contributions to
any order;

\item{(3)} the prepotential in the limit of large hypermultiplet 
mass $m$ (as well as large gauge scalar expectation
value and small gauge coupling) correctly reproduces the 
prepotentials for $\N=2$ super Yang Mills theory 
with any number of hypermultiplets in the fundamental representation of the
gauge group.

\medskip

The fundamental problem in Seiberg-Witten theory
is to determine the Seiberg-Witten curves and differentials, 
corresponding to an $\N=2$ supersymmetric gauge theory with 
arbitrary gauge algebra $\G$, and a massive hypermultiplet in
an arbitrary representation $R$ of $\G$, subject to
the constraint of asymptotic freedom or conformal invariance.
With the correspondence between Seiberg-Witten 
curves and the spectral curves of classical mechanics 
integrable systems [3], this problem is equivalent to 
determining a general integrable system, associated 
with the Lie algebra $\G$ and the representation $R$.

\medskip

The $\N=2$ theory for arbitrary gauge algebra $\G$ and
with one massive hypermultiplet in the adjoint representation 
was one such outstanding case when $\G \not= SU(N)$.
Actually, as discussed previously, upon taking suitable 
limits, this theory contains a very large number of
models with smaller hypermultiplet representations $R$,
and in this sense has a universal aspect. It appeared
difficult to generalize directly the Donagi-Witten
construction of Hitchin systems
to arbitrary $\G$, and it was thus natural 
to seek this generalization directly amongst the elliptic Calogero-Moser
integrable systems.
It has been known now for a long time,
thanks to the work of Olshanetsky and Perelomov [6],
that Calogero-Moser systems can be defined for any simple
Lie algebra.
Olshanetsky and Perelomov also showed that
the Calogero-Moser systems for {\it classical} Lie
algebras were integrable,
although the existence of a spectral curve
(or Lax pair with spectral parameter) as well as the case
of exceptional Lie algebras remained open.
Thus several immediate questions are:

\medskip

$\bullet$ Does the elliptic Calogero-Moser system for general
Lie algebra $\G$ admit a Lax pair with spectral parameter? 

$\bullet$ Does it correspond to the $\N=2$ supersymmetric gauge
theory with gauge algebra $\G$ and a hypermultiplet in
the adjoint representation?

$\bullet$ Can this correspondence be verified in the limiting
cases when the mass $m$ tends to $0$ with the theory acquiring $\N=4$ 
supersymmetry and when $m\rightarrow\infty$, with the hypermultiplet
decoupling in part to smaller representations of $\G$ ? 

\medskip
 
\noindent
The purpose of this paper is to review the solution to these
questions, which were obtained in [7], [8] and [9]. In summary, the
answers can be stated succinctly as follows.

\medskip

$\bullet$ The elliptic Calogero-Moser systems defined
by an arbitrary simple Lie algebra $\G$ do admit Lax
pairs with spectral parameters.

$\bullet$ The correspondence between elliptic $\G$ Calogero-Moser
systems and $\N=2$ supersymmetric $\G$ gauge theories
with matter in the adjoint representation holds directly when
the Lie algebra $\G$ is simply-laced.
When $\G$ is not simply-laced, the correspondence is with 
new integrable models, {\it the
twisted elliptic Calogero-Moser systems}
introduced in [7,8]. 

$\bullet$ The new twisted elliptic Calogero-Moser systems
also admit a Lax pair with spectral parameter [7].

$\bullet$ In the scaling limit $m=Mq^{-{1\over 2}\d}\rightarrow\infty$,
$M$ fixed, the twisted (respectively untwisted) elliptic $\G$
Calogero-Moser
systems tend to the Toda system for
$(\G^{(1)})^{\vee}$ (respectively $\G^{(1)}$)
for $\d={1\over h_{\G}^{\vee}}$ (respectively $\d={1\over h_{\G}}$).
Here $h_{\G}$ and $h_{\G}^{\vee}$ are the Coxeter and the dual Coxeter numbers
of $\G$ [8]. 

\medskip

The remainder of this paper is organized as follows. In \S II, we briefly
review supersymmetric gauge theories, the set-up and basic constructions
of Seiberg-Witten theory. In \S III, we discuss the elliptic Calogero-Moser 
systems introduced by Olshanetsky and Perelomov long ago, and present
the new twisted elliptic Calogero-Moser systems introduced in [7,8]. In \S IV, we
show how these systems tend to the Toda systems in certain limits and 
discuss their integrability properties and Lax pairs with spectral parameter
in \S V. Finally, in \S VI and \S VII, we discuss the Seiberg-Witten solution 
for the $\N=2$ supersymmetric Yang-Mills theories and a massive hypermultiplet 
in the adjoint representation of the gauge algebra for $\G=SU(N)$ and for 
arbitrary $\G$ respectively.

\vfill\eject

\centerline{\bf II. SEIBERG-WITTEN THEORY}

\bigskip

Supersymmetric Yang-Mills theories are ordinary field theories of
scalar, spin 1/2 fermions and gauge fields, with field contents
fitting into representations of the supersymmetry algebra and 
with certain special relations between the gauge, Yukawa and Higgs 
self-couplings. For each of $\N=1,\ 2,\ 4$, there is a gauge 
multiplet (g) in the adjoint representation of the gauge algebra $\G$
and for $\N=1,\ 2$ there are matter multiplets (m) in an arbitrary
representation $R$ of $\G$.

\medskip

\noindent
{\bf a) Supersymmetry multiplets}

\medskip

\item{(1)} For $\N=1$, we have 

\itemitem{(g)} the gauge multiplet $(A_\mu, \lambda)$ containing 
a gauge field $A_\mu$ and a Majorana fermion~$\lambda$;

\itemitem{(m)} the chiral multiplet $(\varphi, \psi)$ containing 
a complex scalar $\varphi$ and a chiral fermion $\psi$.

\item{(2)} For $\N=2$, we have

\itemitem{(g)} the gauge multiplet $(A_\mu, \lambda _\pm, \phi)$
containing a gauge field $A_\mu$, a Dirac fermion $\lambda _\pm$ and
a complex scalar $\phi$, which we shall often refer to as the gauge scalar.
Under an $\N=1$ supersymmetry subalgebra, the $\N=2$ gauge multiplet is the
direct sum of the $\N=1$ gauge multiplet and an $\N=1$ chiral multiplet in the
adjoint representation of $\G$.

\itemitem{(m)} the hypermultiplet $(\psi_\pm, H_\pm)$ contains 
a Dirac fermion $\psi _\pm$ and two complex scalars $H_\pm$. Under
an $\N=1$ subalgebra, this multiplet is the sum of one left and
one right $\N=1$ chiral multiplets.

\item{(3)} For $\N=4$, we have

\itemitem{(g)} the gauge multiplet $(A_\mu,\lambda _\alpha, \phi _I)$,
containing a gauge field $A_\mu$, four Majorana spinors $\lambda _\alpha$, 
$\alpha =1,\cdots,4$ and six real scalars $\phi _I$, $I=1,\cdots ,6$.
Under an $\N=2$ subalgebra, the multiplet is the sum of an $\N=2$
gauge multiplet and an $\N=2$ hypermultiplet in the adjoint representation of
the gauge algebra $\G$.

\itemitem{(m)} there is no matter multiplet for $\N=4$.

\bigskip

\noindent
{\bf b) Supersymmetric Lagrangians}

\medskip

For the study of Seiberg-Witten theory, we shall need both
the $\N=2$ supersymmetric microscopic (renormalizable) Lagrangian
as well as $\N=2$ supersymmetric effective Lagrangians. Both
types may be viewed as general Lagrangians involving the multiplets
given above, but with the restriction that only terms are retained
with at most two derivatives on any term involving boson fields, and one 
derivative on any term involving fermion fields. This is the usual
approximation made when dealing with effective low energy theories, and
also happens to be one of the criteria for renormalizability. These effective
Lagrangians are always polynomial in the gauge and fermion fields, but
depend upon the various scalar fields through possibly general functions.
Supersymmetry imposes certain holomorphicity conditions on some
of these functions, a property fundamental in the Seiberg-Witten analysis.
Henceforth, we restrict to considering only such Lagrangians.

\medskip

For $\N=1$ supersymmetric theories with gauge multiplet $(A_\mu ^a, \lambda 
^a)$, $a=1,\cdots , \dim \G$ and chiral multiplets $(\varphi ^i, \psi ^i)$, 
$i=1,\cdots ,N_f$, the key parts of the most general Lagrangian are
given by the kinetic terms type and potential terms for the fields (all other
terms such as Yukawa couplings are omitted, as we shall not need their form) 
$$
\eqalign{
{\cal L}
= &-g_{i\bar j} \bigl [ D_\mu \varphi ^i D^\mu \varphi ^{\bar j}
+ i\bar \psi ^{\bar j} \bar \sigma ^\mu D_\mu \psi ^i \bigr ] -\half
g^{i\bar j} {\partial W(\varphi) \over \partial \varphi ^i}
{\partial \bar W(\bar \varphi) \over \partial \bar \varphi ^{\bar j}} \cr
& -\half \tau _{ab} (\varphi) \bigl [
  {i \over 16} F_{\mu \nu} ^a F^{\mu \nu a} 
  -{1 \over 16} F_{\mu \nu} ^a \tilde F^{\mu \nu a} 
  + \bar \lambda ^b \bar \sigma ^\mu D_\mu \lambda ^a \bigr ] + c.c. + \cdots
\cr}
\eqno (2.1)
$$
Here $g_{i\bar j} = \partial _i \partial _{\bar j} K(\varphi, \bar \varphi)$
is the K\"ahler metric on the scalar fields, $D_\mu$ are suitable covariant 
derivatives with respect to the gauge field (and the K\"ahler connection for 
$D_\mu $ on fermions), and $F_{\mu \nu}$ is the field strength of $A_\mu$.
The superpotential $W(\varphi)$ and the gauge coupling field $\tau 
_{ab}(\varphi)$ are constrained by $\N=1$ supersymmetry to be {\it complex 
analytic} functions of $\varphi$. For $\N=1$ supersymmetric theories,
it is very convenient to derive the above results from an $\N=1$ superfield
formulation, in which the complex analyticity of $W$ and $\tau _{ab}$ 
emerges from the fact that these functions arise in $F$-terms, while the 
K\"ahler potential comes from a $D$-term. In $F$-terms, only superfields of one 
chirality enter; since a chiral fermion is in the same multiplet as the complex 
scalar field $\varphi$, but not $\bar \varphi$, all $\varphi$-dependence 
emerging from $F$-terms is inherently complex analytic. With a generalization to $\N=2$ and 
$\N=4$ in mind, where no convenient off-shell superfield formulation
is available, we prefer here to use component language throughout.

\medskip

For $\N=2$ supersymmetric theories, the gauge multiplet consists of 
an $\N=1$ gauge multiplet and an $\N=1$ chiral multiplet in the adjoint
representation of the gauge algebra. Thus, part of the components of
the chiral field $(\varphi ^i, \psi ^i)$ are in the adjoint representation,
and we shall denote that part by $(\phi ^a, \psi ^a)$, with the index $a$ 
running through the adjoint representation. (The remaining components make up
hypermultiplets.) Since the adjoint representation is real,  there is no
distinction between $a$ and $\bar a$. We shall concentrate on that  part of the
Lagrangian (2.1) which involves only the $\N=2$ vector multiplet  fields. 

\medskip

Enforcing $\N=2$ supersymmetry on the $\N=1$ Lagrangian (2.1)
for the vector multiplet is not so easy. However, it is straightforward
to enforce some necessary conditions. The $\N=2$ supersymmetry algebra
is invariant under an $SU(2)_R$ group which rotates the two independent
supercharges into one another, and thus rotates the two spinors in the $\N=2$
gauge multiplet into one another as well. In the $\N=1$ language used in (2.1), 
these two spinors are $\psi ^a$ and $\lambda ^a$. $\N=2$ supersymmetry requires 
invariance under $SU(2)_R$, and thus invariance of the Lagrangian under this 
symmetry. Invariance of the kinetic terms for $\lambda$ and $\psi$ in (2.1)
immediately yields a relation between the K\"ahler metric and the gauge 
coupling 
function
$$
{\partial ^2 K (\phi, \bar \phi) \over \partial \phi ^a \partial \bar \phi ^b}
 = {\rm Im}\, \tau _{ab} (\phi)
\eqno (2.2)
$$
Since $\tau _{ab}(\phi)$ is a complex analytic function of $\phi$, the partial 
derivative of (2.2) with respect to $\phi ^a$ is complex analytic, and thus
$$
{\partial ^2 K (\phi, \bar \phi) \over \partial \phi ^a \partial \phi ^c}
 = T_{abc} (\phi) \bar \phi ^b
\eqno (2.3)
$$
for some complex analytic function $T_{abc}$ of $\phi$. The most general 
solution to (2.3) is very easily obtained by integrating up twice, and may be 
expressed in terms of a single complex analytic function $\F$, called the 
superpotential. In terms of $\F$, the quantities $\tau$ and $K$ are given by
$$
\eqalignno{
\tau _{ab}(\phi) & =
 {\partial ^2 \F (\phi) \over \partial \phi ^a \partial \phi ^b}
& (2.4a) \cr
K(\phi, \bar \phi) & =
 -{i \over 2} \bar \phi ^b {\partial \F \over \partial \phi ^b}
+{i \over 2} \phi ^b {\partial \bar \F \over \partial \bar \phi ^b}
& (2.4b) \cr}
$$
This restricted form of the $\N=2$ effective action is closely related with 
special geometry. 

\medskip

Imposing $SU(2)_R$-symmetry on the Yukawa couplings requires that the 
superpotential for the gauge scalars be similarly restricted. The resulting 
expressions are rather complicated, and we shall give below only the special 
cases needed for our analysis. Analogous conditions are required upon inclusion 
of hypermultiplets, but we shall not give those here.
Once these necessary conditions arising from $SU(2)_R$ invariance have been 
imposed, it may in fact be shown that the Lagrangian obtained in this way is 
indeed $\N=2$ supersymmetric [10].

\medskip

\noindent
{\bf c) The Set-Up for Seiberg-Witten Theory}

\medskip

The starting point for Seiberg-Witten theory is an $\N=2$ supersymmetric 
Yang-Mills theory with gauge algebra $\G$ and hypermultiplets in a 
representation $R$ of $\G$ with masses $m_j$. The microscopic Lagrangian is 
completely fixed by $\N=2$ supersymmetry in terms of the gauge coupling $g$ and 
the instanton angle $\theta$, and is given by
$$
{\cal L} = {1 \over 4 g^2} F_{\mu \nu}^a F^{\mu \nu a}
 + {\theta \over 32 \pi ^2} F_{\mu \nu}^a \tilde F^{\mu \nu a}
 + D_\mu \bar \phi D^\mu \phi + \tr [\bar \phi , \phi ]^2 + \cdots
 \eqno (2.5)
$$
where we have neglected hypermultiplet and fermion terms.

\medskip

The low energy effective theory corresponding to this model can be analyzed
by studying first the structure of the vacuum. $\N=2$ supersymmetric vacuum 
states can occur whenever the vacuum energy is exactly zero. Since the energy 
is 
always positive in a supersymmetric theory, we are guaranteed that any zero 
energy solution is a vacuum. This is the case here for vanishing gauge fields
and  constant gauge 
scalar fields $\phi$ for which the potential energy term also vanishes.
The potential energy vanishes if and only if $[\bar \phi, \phi]=0$,  
a condition equivalent to the vacuum expectation value of $\phi$ 
being a linear combination of the Cartan generators of the gauge algebra $\G$,
$$
< \phi > = \sum _{j=1} ^n a_j h_j
\qquad \qquad
n={\rm rank} ~\G
\eqno (2.6)
$$
Here, the complex parameters $a_j$ are usually referred to as the quantum 
moduli, or also as the quantum order parameters of the $\N=2$ vacua. 

\medskip

For generic values of the parameters $a_j$, the $\G$-gauge symmetry will be 
broken down to $U(1)^n/{\rm Weyl}(\G)$, and the low energy theory is
that of $n$ different Coulomb fields, up to global identifications by ${\rm
Weyl}(\G)$. Since
$\N=2$ supersymmetry is unbroken in  any of these vacua, the low energy effective
Lagrangian will have to be  invariant under $\N=2$ supersymmetry. But, we have
already given a description  of all such effective actions before, in terms of a
complex analytic  superpotential $\F (\phi)$. In the case of $n$ different
$U(1)$ gauge fields,  this effective Lagrangian is particularly simple, and we
have
$$
{\cal L} _{\rm effective} 
= {1 \over 4}{\rm Im} (\tau _{ij}) F_{\mu \nu} ^i F^{\mu \nu j}
+ {1 \over 4} {\rm Re} (\tau _{ij}) F_{\mu \nu} ^i \tilde F^{\mu \nu j}
+ \partial _\mu \bar \phi ^j \partial ^\mu \phi _{Dj} + {\rm fermions}
\eqno (2.7)
$$
Here, the dual gauge scalar $\phi _D$ and the gauge coupling function $\tau 
_{ij}$ are both given in terms of the prepotential $\F$
$$
\phi _{Dj} = {\partial \F(\phi) \over \partial \phi _j}
\qquad
\tau _{ij} = {\partial ^2 \F (\phi) \over \partial \phi _i \partial \phi _j}
\eqno (2.8)
$$
The form of the effective Lagrangian (2.7) is the same for any of the values
of the complex  moduli of $\N=2$ vacua, with the understanding that the
fields $\phi  _j$ take on the expectation value $<\phi _j>=a_j$. Since the
prepotential $\F(\phi)$ is a function of the fields $\phi$ only, but not of
derivatives of $\phi$, the prepotential will be completely determined by its
values on the  vacuum expectation values of the field, namely by its values on
the quantum order parameters $a_j$.

\medskip

\noindent
{\bf d) The Seiberg-Witten Solution}

\medskip

The object of Seiberg-Witten theory is the determination of the prepotential 
$\F (a_j)$, from which the entire low energy effective action will be known. 
This is achieved by exploiting the physical conditions satisfied by $\F$ [1],

\medskip 

\item{(1)} $\F(a_j)$ is complex analytic in $a_j$ in view of $\N=2$ 
supersymmetry, as shown in b) above.

\item{(2)} The matrix ${\rm Im}\ \tau _{ij} = {\rm Im} \partial _i \partial _j 
\F$ is positive definite, since by (2.7), it coincides with the metric on the 
kinetic terms for the gauge fields $A_j$.

\item{(3)} The large $a_j$ behavior is known from perturbative quantum field 
theory calculations and asymptotic freedom, and is given by
$\F(a) \sim (a_i-a_j)^2 \ln (a_i-a_j)^2$. 

\medskip

More precisely, for gauge algebra $\G$ and hypermultiplets in the representation
$R$ of $\G$, $\F(a)$ is of the form
$$
\eqalign{\F(a)=&\F^{\rm class}(a)  +\sum_{d=1}^{\infty}\F_d (a) \Lambda^{(2h_{\G}^{\vee}-I(R))d}
\cr
&-{1\over 8\pi i}
\big[\sum_{\alpha\in {\cal R}(\G)}(\alpha\cdot a)^2{\rm ln}\,{(\alpha\cdot a)^2
\over \Lambda^2}
-\sum_{\lambda\in W(R)}(\lambda\cdot a+m)^2{\rm ln}\,{(\lambda\cdot a+m)^2
\over\Lambda^2}
\big].\cr}
\eqno(2.9)
$$
Here $\Lambda$ is a dynamically generated scale introduced by
renormalization, $h_{\G}^{\vee}$ is the quadratic Casimir of $\G$ (equal to 
the dual Coxeter number),
$I(R)$ is the Dynkin index of the representation $R$,
and ${\cal R}(\G)$ and $W(R)$ denote respectively
the roots of $\G$ and the weights of the representation $R$.
The terms on the right hand side of (2.9)
represent respectively the classical prepotential, the
one-loop perturbative corrections (higher loops do not
contribute in view of non-renormalization theorems),
and the instanton corrections $\F^{(d)}=\F_d\Lambda^{(2h_{\G}^{\vee}-I(R))d}$
of all orders $d$.
In general, it is prohibitively difficult
to determine the coefficients $\F_d$ from field theory methods.
For conformally invariant theories, the expansion (2.9)
is replaced by a similar one where the dynamical scale
$\Lambda$ is replaced by a modular invariant $q=e^{2\pi i\tau}$
(see e.g. (6.2) and (6.7) below).

\medskip

\noindent
As a result of the requirements (1) and (2), it follows immediately that $\F$ 
cannot be a single-valued function of the $a_j$. For if it were, ${\rm Im} \ 
\tau _{ij}$ would be both harmonic and bounded from below, which would imply 
that it must be independent of $a_j$. But, from (3), we know that $\tau _{ij}$ 
is not constant at large $a_j$. And indeed, from (3) again, it is clear that 
neither $\F$ nor $\tau_{ij}$ are single valued functions of the $a_j$.

\medskip

As is clear from the large $a_j$ behavior $\tau_{ij} (a) \sim \ln (a_i - a_j)$, 
one of the ways in which $\tau _{ij}(a)$ is multiple valued is by shifts of
any of the matrix elements by an integer. This ambiguity does not affect the 
physics of the low energy effective action (2.7), because the constant shifts 
in ${\rm Re} (\tau _{ij})$ are like the shifts of the instanton angle $\theta$ 
by $2\pi$ times an integer and not observable. A more complicated 
multiple-valuedness consists in taking $\tau \to - \tau ^{-1}$, and corresponds 
to electric-magnetic duality, as shown by Seiberg and Witten [1]. The
combination  of these two types of transformations produces the full duality
group $SL(2n,{\bf Z})$ of monodromies of $\tau$.

\medskip

A natural setting in which the above monodromy problem may be solved is 
provided by families of Riemann surfaces, called the Seiberg Witten curves, 
denoted by $\Gamma$. Indeed, letting the quantum moduli $a_j$ correspond to 
moduli of the Riemann surfaces, there is automatically a complex analytic 
period matrix, whose imaginary part is positive definite, and whose monodromy 
group corresponds to the modular group of the surface.
For $\G=SU(2)$ gauge group and no hypermultiplets for example, the 
Seiberg-Witten curve is a of genus 1, and may be represented as a double 
sheeted cover of the complex plane, $\Gamma (u) = \{ (x,y); \ y^2 = 
(x-\Lambda)(x+\Lambda ) (x-u)\}$. Here $u$ is an auxiliary parameter, which
will be related to the quantum modulus $a$, and $\Lambda$ is the renormalization 
scale. We shall choose the branch cut between the points $x=\pm \Lambda$. The 
quantum modulus and prepotential are then given by
$$
a(u) = {1\over 2\pi i}\oint  _A (x-u) {dx \over y}
\qquad \qquad
a_D(u) = {\partial \F (a) \over \partial a} = {1\over 2\pi i}\oint _B (x-u) {dx \over y}
\eqno (2.10)
$$
where the $A$-cycle may be chosen around the branch cut between $\pm \Lambda$ 
and the $B$-cycle between the branch points $+\Lambda $ and $u$. As $u\to \pm 
\Lambda$, the elliptic curve produces a singularity which physically is 
interpreted as caused by the vanishing of the mass of a magnetic monopole or 
dyon.

\medskip

Starting from the Seiberg-Witten solution for gauge group $\G=SU(2)$, one may 
abstract 
the general set-up of the Seiberg-Witten solution, expected for arbitrary gauge 
algebra $\G$ with rank $n$ and general hypermultiplet representation. The 
ingredients are

\medskip

\item{(1)} The Seiberg-Witten curve is a family of Riemann surfaces $\Gamma 
(u_1, \cdots , u_n)$ dependent on $n$ auxiliary complex parameters $u_j$, which
are related to the quantum moduli $a_j$. The Seiberg-Witten curve will also 
depend upon the gauge coupling $g$ and $\theta$-angle and on the hypermultiplet
masses $m_k$.

\item{(2)} The Seiberg-Witten meromorphic differential 1-form $d\lambda$ on 
$\Gamma$, whose residues are linear in the hypermultiplet masses $m_k$. Since 
the hypermultiplet masses receive no quantum corrections as $a_j$ varies,
the derivatives $\partial (d\lambda)/\partial a_j$ are holomorphic 1-forms.

\item{(3)} The quantum moduli and the prepotential are given by
$$
a_j = {1\over 2\pi i}\oint _{A_j} d\lambda \qquad \qquad
a_{Dj} = {\partial \F \over \partial a_j} = {1\over 2\pi i}\oint _{B_j} d \lambda
\eqno (2.11)
$$

Shortly after the initial work of Seiberg and Witten, the curves and 
differentials for general $SU(N)$, with and without hypermultiplets
in the fundamental representation were proposed, as well as generalizations to 
the gauge groups $SO(N)$ and $Sp(N)$ [11]. See also [12-14]. Use was made of the $R$-charge assignments of the fields, the singularity structure of the degenerations of  the Seiberg-Witten curve, and much educated guess work.

\bigskip
\bigbreak

\centerline{\bf III. TWISTED AND UNTWISTED CALOGERO-MOSER SYSTEMS}

\bigskip

\noindent
{\bf a) The $SU(N)$ Elliptic Calogero-Moser System}

The original elliptic Calogero-Moser
system is the system defined by the Hamiltonian
$$
H(x,p)={1\over 2}\sum_{i=1}^Np_i^2-{1\over 2}m^2
\sum_{i\not=j}\wp(x_i-x_j)
\eqno(3.1)
$$
Here $m$ is a mass parameter, and $\wp(z)$ is the Weierstrass $\wp$-function,
defined on a torus ${\bf C}/(2\omega_1{\bf Z}+2\omega_2{\bf Z})$. As usual, we
denote by $\tau=\omega_2/\omega_1$ the moduli of the torus, and set $q=e^{2\pi 
i\tau}$. The well-known trigonometric and rational limits with respective
potentials 
$$
-{1\over 2}m^2\sum_{i\not=j}
{1\over 4\,{\rm sh}^2\,({x_i-x_j\over 2})}
\qquad
{\rm and}
\qquad 
-{1\over 2}m^2\sum_{i\not=j}{1\over (x_i-x_j)^2}
$$ 
arise in the limits $\omega_1=-i\pi,\omega_2\rightarrow\infty$
and $\omega_1,\omega_2\rightarrow\infty$.
All these systems have been shown to be completely integrable
in the sense of Liouville, i.e. they all admit a complete set of integrals 
of motion which are in involution [15-17]. For a recent review of some applications of these models see [18].

\bigskip

Our considerations require however a notion of integrability which
is in some sense more stringent, namely the existence of a Lax pair $L(z)$,
$M(z)$ with spectral parameter $z$. Such a Lax pair was obtained by Krichever
[19] in 1980. He showed that the Hamiltonian system (3.1) is
equivalent to the Lax equation $\dot L(z)=[L(z),M(z)]$,
with $L(z)$ and $M(z)$ given by the following $N\times N$ matrices
$$
\eqalignno{
L_{ij}(z)=&p_i\d_{ij}-m(1-\d_{ij})\P(x_i-x_j,z)\cr
M_{ij}(z)=&m\d_{ij}\sum_{k\not= i}\wp(x_i-x_k)-m(1-\d_{ij})\P'(x_i-x_j,z).
&(3.2)\cr}
$$ 
The function $\P(x,z)$ is defined by
$$
\P(x,z)={\sigma(z-x)\over
\sigma(z)\sigma(x)}e^{x\zeta(z)},
\eqno(3.3)
$$
where $\sigma(z)$, $\zeta(z)$ are the usual Weierstrass $\sigma$ and $\zeta$
functions on the torus ${\bf C}/(2\omega_1{\bf Z}+2\omega_2{\bf  Z})$.
The function $\P(x,z)$ satisfies the key functional equation
$$
\Phi(x,z)\P'(y,z)
-\P(y,z)\P'(x,z)
=(\wp(x)-\wp(y))\P(x+y,z).
\eqno(3.4)
$$ 
It is well-known that functional equations of this form 
are required for the Hamilton
equations of motion to be equivalent to the Lax equation
$\dot L(z)=[L(z),M(z)]$ with a Lax pair of the form (3.2).
Often, solutions had been obtained under
additional parity assumptions in $x$ (and $y$), which prevent
the existence of a spectral parameter. 
The solution $\P(x,z)$ with spectral parameter $z$
is obtained by dropping such parity assumptions
for general $z$. It is a relatively recent result
of Braden and Buchstaber [20] that, conversely,
general functional equations of the form (3.4) essentially
determine $\P(x,z)$.

\bigskip

\noindent
{\bf b) Calogero-Moser Systems defined by Lie Algebras}

As Olshanetsky and Perelomov [6] realized very early on,
the Hamiltonian system (3.1) is only one
example of a whole series of Hamiltonian systems associated
with each simple Lie algebra.
More precisely, given any simple Lie algebra $\G$,
Olshanetsky and Perelomov [6] introduced the system
with Hamiltonian
$$
H(x,p)
={1\over 2}\sum_{i=1}^rp_i^2
-{1\over 2}
\sum_{\a\in{\cal R}(\G)}
m_{|\a|}^2\wp(\a\cdot x),
\eqno(3.5)
$$
where $r$ is the rank of $\G$, ${\cal R}(\G)$
denotes the set of roots of $\G$, and the $m_{|\a|}$ are mass parameters.
To preserve the invariance of the Hamiltonian (3.5)
under the Weyl group, the parameters $m_{|\a|}$ depend only
on the orbit $|\a|$ of the root $\a$, and not on the root $\a$ itself.
In the case of $A_{N-1}= SU(N)$, it is common practice to use $N$ pairs of
dynamical variables $(x_i,p_i)$, since the roots of $A_{N-1}$
lie conveniently on a hyperplane in ${\bf C}^N$.
The dynamics of the system are unaffected if we shift
all $x_i$ by a constant, and the number of degrees of freedom
is effectively $N-1=r$. Now the roots of $SU(N)$ are given
by $\a=e_i-e_j$, $1\leq i,j\leq N$, $i\not=j$. Thus
we recognize the original elliptic Calogero-Moser system
as the special case of (3.5) corresponding to $A_{N-1}$.
As in the original case, the elliptic systems (3.5)
admit rational and trigonometric limits.
Olshanetsky and Perelomov succeeded in constructing a
Lax pair for all these systems in the case of classical
Lie algebras, albeit without spectral parameter [6].

\bigskip
\noindent
{\bf c) Twisted Calogero-Moser Systems defined by Lie Algebras}

It turns out that the Hamiltonian systems (3.5) are not the
only natural extensions of the basic elliptic Calogero-Moser
system. 
A subtlety arises for simple Lie algebras $\G$ which are not
simply-laced, i.e., algebras which admit roots of uneven
length. This is the case for the algebras $B_n$, $C_n$, $G_2$,
and $F_4$ in Cartan's classification.
For these algebras, the following {\it twisted} elliptic
Calogero-Moser systems were introduced by the authors in [7,8]
$$
H_{\G}^{{\rm twisted}}
=
{1\over 2}\sum_{i=1}^rp_i^2
-{1\over 2}
\sum_{\a\in{\cal R}(\G)}
m_{|\a|}^2
\wp_{\nu(\a)}(\a\cdot x).
\eqno(3.6)
$$
Here the function $\nu(\a)$ depends only on the length of the root $\a$.
If $\G$ is simply-laced, we set $\nu(\a)=1$ identically. Otherwise,
for $\G$ non simply-laced, we set $\nu(\a)=1$ when $\a$ is a long root,
$\nu(\a)=2$ when $\a$ is a short root and $\G$ is one of the
algebras $B_n$, $C_n$, or $F_4$, and $\nu(\a)=3$ when $\a$ is a short root
and $\G=G_2$. The {\it twisted} Weierstrass function $\wp_{\nu}(z)$
is defined by
$$
\wp_{\nu}
(z)
=\sum_{\sigma=0}^{\nu-1}
\wp(z+2\omega_a{\sigma\over\nu}),
\eqno(3.7)
$$
where $\omega_a$ is any of the half-periods $\omega_1$,
$\omega_2$, or $\omega_1+\omega_2$.
Thus the twisted and untwisted Calogero-Moser systems coincide
for $\G$ simply laced. 
The original motivation for twisted
Calogero-Moser systems was based on
their scaling limits
(which will be discussed in the next section) [7,8].
Another motivation based on the symmetries of Dynkin diagrams
was proposed subsequently by Bordner, Sasaki, and Takasaki [21].

\bigskip
\bigbreak

\centerline{\bf IV. SCALING LIMITS OF CALOGERO-MOSER SYSTEMS}

\bigskip

\noindent
{\bf a) Results of Inozemtsev for $A_{N-1}$}

For the standard elliptic Calogero-Moser systems
corresponding to $A_{N-1}$, Inozemtsev [22] has shown in the
1980's that in the scaling limit
$$
\eqalignno{m&=Mq^{-{1\over 2N}},\ \ \ \ q\rightarrow0&(4.1)\cr
x_i&=X_i-2\omega_2{i\over N},
\
1\leq i\leq N &(4.2)\cr}
$$
where $M$ is kept fixed,
the elliptic $A_{N-1}$ Calogero-Moser Hamiltonian tends to
the following Hamiltonian 
$$
H_{{\rm Toda}}
={1\over 2}\sum_{i=1}^Np_i^2
-{1\over 2}\big(\sum_{i=1}^{N-1}e^{X_{i+1}-X_i}+e^{X_1-X_N}\big)
\eqno(4.3)
$$
The roots $e_i-e_{i+1}$, $1\leq i\leq N-1$, and $e_N-e_1$
can be recognized as the simple roots of the
affine algebra $A_{N-1}^{(1)}$.
(For basic facts on affine algebras, we refer to [23]).
Thus (4.3) can be recognized as the Hamiltonian of the Toda system
defined by $A_{N-1}^{(1)}$.

\bigskip
\noindent
{\bf b) Scaling Limits based on the Coxeter Number}

The key feature of the above scaling limit is the collapse
of the sum over the entire root lattice of $A_{N-1}$
in the Calogero-Moser Hamiltonian to the
sum over only simple roots in the Toda Hamiltonian for the
Kac-Moody algebra $A_{N-1}^{(1)}$.
Our task is to extend this mechanism to general Lie algebras.
For this, we consider the following generalization  of the preceding scaling
limit
$$
\eqalignno{
m&=Mq^{-{1\over 2}\d},&(4.4)\cr
x&=X-2\omega_2\d\rho^{\vee},&(4.5)\cr}
$$
Here $x=(x_i)$, $X=(X_i)$ and $\rho^{\vee}$
are $r$-dimensional vectors.
The vector $x$ is the dynamical
variable of the Calogero-Moser system.
The parameters $\d$ and $\rho^{\vee}$
depend on the algebra $\G$ and are yet to be chosen.
As for $M$ and $X$, they have the same interpretation as
earlier, namely as respectively the mass parameter
and the dynamical variables of the limiting system.
Setting $\omega_1=-i\pi$,
the contribution of each root $\a$ to the Calogero-Moser
potential can be expressed as
$$
m^2\wp(\a\cdot x)
=
{1\over 2}M^2
\sum_{n=-\infty}^{\infty}
{e^{2\d\omega_2}\over
{\rm ch}(\a\cdot x-2n\omega_2)-1}
\eqno(4.6)
$$
It suffices to consider positive roots $\a$.
We shall also assume that $0\leq \d\,\a\cdot\rho^{\vee}
\leq 1$. The contributions of the $n=0$ and $n=-1$
summands in (4.6) are proportional
to $e^{2\omega_2(\d-\d\,\a\cdot\rho^{\vee})}$
and $e^{2\omega_2(\d-1+\d\,\a\cdot\rho^{\vee})}$
respectively.
Thus the existence of a finite scaling limit requires
that
$$
\d\,\leq\d\,\a\cdot\rho^{\vee}\leq 1-\d.
\eqno(4.7)
$$
Let $\a_i$, $1\leq i\leq r$ be a basis of simple roots
for $\G$. If we want all simple roots $\a_i$
to survive in the limit, we must require that
$$
\a_i\cdot\rho^{\vee}=1,\ \
1\leq i\leq r.
\eqno(4.8)
$$
This condition characterizes the vector $\rho^{\vee}$
as the {\it level vector}.
Next, the second condition in (3.7)
can be rewritten as $\d\{1+max_{\a}\,(\a\cdot\rho^{\vee})\}
\leq 1$. But
$$
h_{\G}=1+max_{\a}\,(\a\cdot\rho^{\vee})
\eqno(4.9)
$$
is precisely the Coxeter number of $\G$,
and we must have $\d\leq {1\over h_{\G}}$.
Thus when $\d<{1\over h_{\G}}$,
the contributions of all the roots except
for the simple roots of $\G$ tend to $0$.
On the other hand, when $\d={1\over h_{\G}}$,
the highest root $\a_0$ realizing the maximum over
$\a$ in (4.9) survives.
Since $-\a_0$ is the additional
simple root for the affine Lie algebra
$\G^{(1)}$, we arrive in this way at the following theorem,
which was proved in [8]

\bigskip
\noindent
{\bf Theorem 1}.
{\it Under the limit (4.4-4.5), with $\d={1\over h_{\G}}$,
and $\rho^{\vee}$ given by the
level vector,
the Hamiltonian of the elliptic Calogero-Moser system
for the simple Lie algebra $\G$
tends to the Hamiltonian of the Toda system
for the affine Lie algebra $\G^{(1)}$.}

\bigskip

\noindent
{\bf (c) Scaling Limit based on the Dual Coxeter Number}

If the Seiberg-Witten spectral curve of the $\N=2$
supersymmetric gauge theory with a hypermultiplet in
the adjoint representation is to be realized as
the spectral curve for a Calogero-Moser system,
the parameter $m$ in the Calogero-Moser system
should correspond to the mass of the hypermultiplet.
In the gauge theory, the dependence of the coupling
constant on the mass $m$ is given by
$$
\tau={i\over 2\pi}h_{\G}^{\vee}{\rm ln}\,{m^2\over M^2}
\qquad
\Longleftrightarrow
\qquad
m=Mq^{-{1\over 2h_{\G}^{\vee}}}
\eqno(4.10)
$$
where $h_{\G}^{\vee}$ is the quadratic Casimir of the
Lie algebra $\G$. This shows that the correct physical
limit, expressing the decoupling of the hypermultiplet
as it becomes infinitely massive,
is given by (3.4), but with $\d={1\over h_{\G}^{\vee}}$.
To establish a closer parallel with our preceding discussion,
we recall that the quadratic Casimir $h_{\G}^{\vee}$
coincides with the {\it dual Coxeter number} of $\G$,
defined by
$$
h_{\G}^{\vee}=1+max_{\a}\,(\a^{\vee}\cdot\rho),
\eqno(4.11)
$$
where $\a^{\vee}={2\a\over\a^2}$ is the coroot associated
to $\a$, and $\rho={1\over 2}\sum_{\a>0}\a$
is the well-known Weyl vector.

For simply laced Lie algebras $\G$ (ADE algebras),
we have $h_{\G}=h_{\G}^{\vee}$, and the preceding scaling limits
apply. However, for non simply-laced algebras
($B_n$, $C_n$, $G_2$, $F_4$), 
we have $h_{\G}>h_{\G}^{\vee}$,
and our earlier considerations show that the untwisted
elliptic Calogero-Moser Hamiltonians do not tend to
a finite limit under (3.10), $q\to 0$,
$M$ is kept fixed.
This is why the twisted Hamiltonian systems (3.6)
have to be introduced. The twisting produces precisely
to an improvement in the asymptotic behavior
of the potential which allows a finite, non-trivial limit.
More precisely,
we can write
$$
m^2\wp_{\nu}(x)
=
{\nu^2\over 2}
\sum_{n=-\infty}^{\infty}
{m^2\over {\rm ch}\,\nu(x-2n\omega_2)-1}.
\eqno(4.12)
$$
Setting $x=X-2\omega_2\d^{\vee}\rho$, we
obtain the following asymptotics
$$
m^2\wp_{\nu}(x)
=\nu^2M^2
\cases{e^{-2\omega_2(\d^{\vee}\a^{\vee}\cdot\rho-\d^{\vee})-\a^{\vee}\cdot 
X}
+e^{-2\omega_2(1-\d^{\vee}\a^{\vee}\cdot\rho-\d^{\vee})+\a^{\vee}\cdot X},
&if $\a$ is long;\cr
e^{-2\omega_2(\d^{\vee}\a^{\vee}\cdot\rho-\d^{\vee})-\a^{\vee}\cdot X},
&if $\a$ is short.\cr}
\eqno(4.13)
$$
This leads to the following theorem [8]

\bigskip
\noindent
{\bf Theorem 2}.
{\it 
Under the limit $x=X+2\omega_2{1\over h_{\G}^{\vee}}\rho$,
$m=Mq^{-{1\over 2h_{\G}^{\vee}}}$,
with $\rho$ the Weyl vector and $q\to 0$,
the Hamiltonian of the twisted elliptic Calogero-Moser system
for the simple Lie algebra $\G$
tends to the Hamiltonian of the Toda system
for the affine Lie algebra $(\G^{(1)})^{\vee}$.}

\bigskip

So far we have discussed only the scaling limits of the Hamiltonians.
However, similar arguments show that the Lax pairs constructed below
also have finite, non-trivial scaling limits whenever
this is the case for the Hamiltonians.
The spectral parameter $z$ should scale as $e^z=Zq^{1\over 2}$,
with $Z$ fixed. The parameter $Z$ can be identified
with the loop group parameter
for the resulting affine Toda system.

\bigskip

\centerline{\bf V. LAX PAIRS FOR CALOGERO-MOSER SYSTEMS}

\bigskip

\noindent
{\bf a) The General Ansatz}

Let the rank of $\G$ be $n$,
and $d$ be its dimension.
Let $\L$ be a representation of $\G$ of dimension $N$,
of weights $\l_I$, $1\leq I\leq N$. Let $u_I\in {\bf C}^N$
be the weights of the fundamental
representation of $GL(N,{\bf C})$. Project
orthogonally the $u_I$'s onto the $\l_I$'s as
$$
su_I=\l_I+u_I,
\ \
\l_I\perp v_J.
\eqno(5.1)
$$
It is easily verified that $s^2$ is the second Dynkin index.
Then
$$
\a_{IJ}=\l_I-\l_J
\eqno(5.2)
$$

\noindent
is a weight of $\L\otimes\L^*$ associated to the
root $u_I-u_J$ of $GL(N,{\bf C})$. The Lax pairs
for both untwisted and twisted Calogero-Moser systems
will be of the form
$$
L=P+X,
\ \
M=D+X,
\eqno(5.3)
$$
where the matrices $P,X,D$, and $Y$ are given by
$$
X=\sum_{I\not=J}C_{IJ}\P_{IJ}(\a_{IJ},z)E_{IJ},
\ \ \
Y=\sum_{I\not=j}C_{IJ}\P'_{IJ}(\a_{IJ},z)E_{IJ}
\eqno(5.4)
$$
and by
$$
P=p\cdot h,
\ \ \ \
D=d\cdot (h\oplus\tilde h)+\Delta.
\eqno(5.5)
$$
Here $h$ is in a Cartan subalgebra ${\cal H}_{\G}$
for $\G$, $\tilde h$
is in the Cartan-Killing orthogonal
complement of ${\cal H}_{\G}$
inside a Cartan subalgebra ${\cal H}$ for $GL(N,{\bf C})$,
and $\Delta$ is in the centralizer of ${\cal H}_{\G}$
in $GL(N,{\bf C})$.
The functions $\P_{IJ}(x,z)$ and the coefficients
$C_{IJ}$ are yet to be determined.
We begin by stating the necessary and sufficient
conditions for the pair $L(z)$, $M(z)$ of (4.1)
to be a Lax pair for the
(twisted or untwisted) Calogero-Moser
systems. For this, it is convenient to
introduce the following notation
$$
\eqalignno{
\P_{IJ}&=\P_{IJ}(\a_{IJ}\cdot x)\cr
\wp_{IJ}'
&=\P_{IJ}(\a_{IJ}\cdot x,z)\P_{JI}'(-\a_{IJ}\cdot x,z)
-\P_{IJ}(-\a_{IJ}\cdot x,z)
\P_{JI}'(\a_{IJ}\cdot x,z).
&(5.6)
\cr}
$$

Then the Lax equation $\dot L(z)
=[L(z),M(z)]$ implies the
Calogero-Moser system if and only
if the following three identities are satisfied
$$
\sum_{I\not=J}C_{IJ}C_{JI}\wp_{IJ}'\a_{IJ}
=
s^2\sum_{\a\in {\cal R}(\G)}
m_{|\a|}^2\wp_{\nu(\a)}(\a\cdot x)
\eqno(5.7)
$$
$$
\sum_{I\not=J}C_{IJ}C_{JI}
\wp_{IJ}'(v_I-v_J)
=0
\eqno(5.8)
$$
$$
\eqalignno{
\sum_{K\not= I,J}
C_{IK}C_{KJ}(\P_{IK}\P_{KJ}'-\P_{IK}'\P_{KJ})
&=
sC_{IJ}\P_{IJ}d\cdot (v_I-v_J)
+
\sum_{K\not= I,J}
\Delta_{IJ}C_{KJ}\P_{KJ}\cr
&\qquad\qquad\qquad
-
\sum_{K\not= I,J}
C_{IK}\P_{IK}\Delta_{KJ}
&(5.9)\cr}
$$
\noindent
The following theorem was established in [7]:

\bigskip

\noindent
{\bf Theorem 3}. {\it A representation $\Lambda$, functions
$\Phi_{IJ}$, and coefficients $C_{IJ}$ with a spectral parameter $z$
satisfying (4.7-4.9) can be found for all twisted and untwisted elliptic
Calogero-Moser systems associated with a simple Lie algebra
$\G$, except possibly in the case of twisted $G_2$.
In the case of $E_8$, we have to assume the existence of 
a $\pm1$ cocycle.}   

\vfill\eject

\noindent
{\bf b) Lax Pairs for Untwisted Calogero-Moser Systems}

We now describe some important
features of the Lax pairs we obtain in this manner.

\bigskip

$\bullet$ In the case of the {\it untwisted} Calogero-Moser systems,
we can choose $\P_{IJ}(x,z)=\P(x,z)$,
$\wp_{IJ}(x)=\wp(x)$ for all $\G$.

\medskip

$\bullet$ $\Delta=0$ for all $\G$, except for $E_8$.

\medskip

$\bullet$ For $A_n$, the Lax pair (3.2-3.3) corresponds
to the choice of the fundamental representation for $\L$.
A different Lax pair can be found by taking
$\L$ to be the antisymmetric representation.
 
\medskip

$\bullet$ For the $BC_n$ system, the Lax pair is
obtained by imbedding $B_n$ in $GL(N,{\bf C})$
with $N=2n+1$. When $z=\omega_a$ (half-period),
the Lax pair obtained this way reduces to
the Lax pair obtained by Olshanetsky and Perelomov [6].

\medskip 

$\bullet$ For the $B_n$ and $D_n$ systems,
additional Lax pairs with spectral parameter
can be found by taking $\L$ to be
the spinor representation.

\medskip

$\bullet$ For $G_2$,
a first Lax pair with spectral parameter can be
obtained by the above construction
with $\L$ chosen to be the ${\bf 7}$ of $G_2$.
A second Lax pair with spectral parameter
can be obtained by restricting the {\bf 8}
of $B_3$ to the ${\bf 7}\oplus{\bf 1}$
of $G_2$.

\medskip

$\bullet$ For $F_4$, a Lax pair can be obtained by
taking $\L$ to be the ${\bf 26}\oplus{\bf 1}$
of $F_4$, viewed as the restriction of
the {\bf 27} of $E_6$ to its $F_4$ subalgebra.

\medskip

$\bullet$ For $E_6$, $\L$ is the {\bf 27} representation.

\medskip

$\bullet$ For $E_7$, $\L$ is the {\bf 56} representation.

\medskip

$\bullet$ For $E_8$, a Lax pair with spectral parameter
can be constructed with $\L$ given by the {\bf 248} representation,
if coefficients $c_{IJ}=\pm 1$ exist
with the following cocycle conditions

$$
\eqalignno{
c(\lambda,\lambda-\d)c(\lambda-\d,\mu)=&
c(\lambda,\mu+\d)c(\mu+\d,\mu)\cr
&{\rm \ when\ \d\cdot\lambda=-\d\cdot\mu=1,
\
\lambda\cdot\mu=0}\cr
c(\l,\mu)c(\l-\d,\mu)=&c(\l,\l-\d)\cr
& {\rm \ when\ \d\cdot\l=\l\cdot\mu=1,
\
\d\cdot\mu=0}\cr
c(\l,\mu)c(\l,\l-\mu)=&
-c(\l-\mu,-\mu)\cr
&{\rm \ when\ \l\cdot\mu=1}.
&(5.10)
\cr}
$$
The matrix $\Delta$ in the Lax pair is then the $8\times 8$ matrix
given by

$$
\eqalignno{
\Delta_{ab}=&
\sum_{\d\cdot\b_a=1\atop \d\cdot\b_b=1}
{m_2\over 2}
\big(c(\b_a,\d)c(\d,\b_b)
+
c(\b_a,\b_a-\d)c(\b_a-\d,\b_b)\big)
\wp(\d\cdot x)\cr
&
-\sum_{\d\cdot\b_a=1\atop \d\cdot\b_b=-1}
{m_2\over 2}
\big(c(\b_a,\d)c(\d,\b_b)
+
c(\b_a,\b_a-\d)c(\b_a-\d,\b_b)\big)
\wp(\d\cdot x)\cr
\Delta_{aa}=&
\sum_{\b_a\cdot\d=1}
m_2\wp(\d\cdot x)
+2m_2\wp(\b_a\cdot x),
&(5.11)
\cr}
$$
where $\b_a$, $1\leq a\leq 8$, is a maximal set of 8 mutually
orthogonal roots.

\medskip

$\bullet$ Explicit expressions for the constants $C_{IJ}$ and the functions $d(x)$,
and thus for the Lax pair are particularly simple when the representation
$\Lambda$ consists of only a single Weyl orbit of weights. This is the case when
$\Lambda$ is either

\item{(1)} the defining representation of $A_n$, $C_n$ or $D_n$;

\item{(2)} any rank $p$ totally anti-symmetric representation of $A_n$;

\item{(3)} an irreducible fundamental spinor representation of $B_n$ or $D_n$;

\item{(4)} the ${\bf 27}$ of $E_6$; the ${\bf 56}$ of $E_7$.

The, the weights $\lambda$ and $\mu$ of $\Lambda$ provide unique labels instead
of $I$ and $J$, and the values of $C_{IJ}=C_{\lambda \mu}$ are given by a
simple formula
$$
C_{\lambda \mu} = \left \{ \matrix{
\sqrt{{\alpha ^2 \over 2}} m_{|\alpha|} &
{\rm when} \ \alpha =\lambda - \mu \ {\rm is \ a \ root} \cr
&\cr
0 & {\rm otherwise} \cr} \right .
$$

The expression for the vector $d$ may be summarized by
$$
s d \cdot u_\lambda = \sum _{\lambda \cdot \delta =1;\ \delta ^2=2}
m_{|\delta |} \wp (\delta \cdot x)
$$
(For $C_n$, the last equation has an additional term, as given in [7].)
In each case, the number of independent couplings $m_{|\alpha|}$ equals the
number of different root lengths.

\bigskip

\noindent
{\bf c) Lax Pairs for Twisted Calogero-Moser Systems}

Recall that the twisted and untwisted Calogero-Moser systems
differ only for non-simply laced Lie algebras, namely
$B_n$, $C_n$, $G_2$ and $F_4$.
These are the only algebras we discuss in
this paragraph.
The construction (4.3-4.9) gives then
Lax pairs for all of them, 
with the possible exception of twisted $G_2$.
Unlike the case of untwisted Lie algebras however,
the functions $\P_{IJ}$ have to be chosen
with care, and differ for each algebra.
More specifically,

\medskip

$\bullet$ For $B_n$, the Lax pair is of dimension $N=2n$,
admits two independent couplings $m_1$ and $m_2$,
and
$$
\P_{IJ}(x,z)
=
\cases{
\P(x,z), &if $I-J\not= 0,\pm n$\cr
\P_2({1\over 2}x,z), &if $I-J=\pm n$\cr}.
\eqno(5.12)
$$
Here a new function $\P_2(x,z)$ is defined by
$$
\P_2({1\over 2}x,z)
={\P({1\over 2}x,z)\P({1\over 2}x+\omega_1,z)
\over
\P(\omega_1,z)}
\eqno(5.13)
$$

$\bullet$ For $C_n$, the Lax pair is of dimension $N=2n+2$,
admits one independent coupling $m_2$,
and
$$
\P_{IJ}(x,z)
=
\P_2(x+\omega_{IJ},z),
$$
where $\omega_{IJ}$ are given by
$$
\omega_{IJ}
=
\cases{0, &if $I\not=J=1,2,\cdots,2n+1$;\cr
\omega_2, &if $1\leq I\leq 2n,\ J=2n+2$;\cr
-\omega_2, &if $1\leq J\leq 2n,\ I=2n+2$.\cr}
\eqno(5.14)
$$

$\bullet$ For $F_4$, the Lax pair is of dimension $N=24$,
two independent couplings $m_1$ and $m_2$,
$$
\P_{\l\mu}(x,z)
=
\cases{\P(x,z), &if $\l\cdot\mu=0$;\cr
\P_1(x,z), &if $\l\cdot\mu={1\over 2}$;\cr
\P_2({1\over 2}x,z), &if $\l\cdot\mu=-1$.\cr}
\eqno(5.15)
$$
where the function $\P_1(x,z)$ is defined by
$$
\P_1(x,z)
=
\P(x,z)
-
e^{\pi i\zeta(z)+\eta_1z}
\P(x+\omega_1,z)
\eqno(5.16)
$$
Here it is more convenient to label
the entries of the Lax pair directly by the weights
$\lambda=\lambda_I$ and
$\mu=\lambda_J$ instead of $I$ and $J$.

$\bullet$ For $G_2$, candidate Lax pairs
can be defined in the {\bf 6} and {\bf 8}
representations of $G_2$, but it is still unknown whether
elliptic functions $\P_{IJ}(x,z)$
exist which satisfy the required identities.

\medskip

We note that recently Lax pairs of root type have been considered [21] which
correspond, in the above Ansatz (5.3-5), to $\Lambda$ equal to the adjoint
representation of $\G$ and the coefficients $C_{IJ}$ vanishing for $I$ or $J$
associated with zero weights. This choice yields another Lax pair for the case
of $E_8$.

\vfill\eject

\centerline{\bf VI. CALOGERO-MOSER AND SEIBERG-WITTEN THEORY FOR SU(N)}

\bigskip

The correspondence between Seiberg-Witten 
theory for $\N=2$ super-Yang-Mills theory with one hypermultiplet in the 
adjoint representation of the gauge algebra, and the elliptic Calogero-Moser 
systems was first established in [5], for the gauge algebra $\G=SU(N)$.
We describe it here in some detail.

\bigskip

All that we shall need here of the elliptic Calogero-Moser system is its Lax
operator $L(z)$, whose $N\times N$ matrix elements are given by
$$
L_{ij}(z) = p_i\delta _{ij} - m(1-\delta _{ij}) \Phi (x_i-x_j,z)
\eqno (6.1)
$$
Notice that the Hamiltonian is simply given in terms of $L$ by $H(x,p) =
\half \tr L(z)^2 + C\wp (z)$ with $C=-\half m^2 N(N-1)$.

\medskip

\noindent
{\bf a) Correspondence of Data}

\medskip

The correspondence between the data of the elliptic Calogero-Moser system and 
those of the Seiberg-Witten theory is as follows. 

\item{(1)} The parameter $m$ in (6.1) is the hypermultiplet mass;

\item{(2)} The gauge coupling $g$ and the $\theta$-angle are related to the 
modulus of the torus $\Sigma ={\bf C} /(2\omega _1 {\bf Z} + 2 \omega _2 {\bf 
Z})$
by 
$$
\tau = {\omega _2 \over \omega _1} = {\theta \over 2 \pi} + {4 \pi i\over g^2}
\, ;
\eqno (6.2)
$$

\item{(3)} The Seiberg-Witten curve $\Gamma$ is the spectral curve of the 
elliptic Calogero-Moser model, defined by
$$
\Gamma = \{ (k,z) \in {\bf C} \times \Sigma, \ \det\bigl (kI-L(z)\bigr )=0\}
\eqno (6.3)
$$
and the Seiberg-Witten 1-form is $d\lambda = k \ dz$. $\Gamma$ is invariant 
under the Weyl group of $SU(N)$. 

\item{(4)} Using the Lax equation $\dot L = [L,M]$, it is clear that the 
spectral curve is independent of time, and can be dependent only upon the 
constants of motion of the Calogero-Moser system, of which there are only $N$.
These integrals of motion may be viewed as parametrized by the quantum moduli
of the Seiberg-Witten system.

\item{(5)} Finally, $d\lambda =kdz$ is meromorphic, with a simple pole on 
each of the $N$ sheets above the point $z=0$ on the base torus. The residue at 
each of these poles is proportional to $m$, as required by the general set-up 
of Seiberg-Witten theory, explained in \S II.

\vfill\eject

\noindent
{\bf b) Four Fundamental Theorems}

\medskip

While the above mappings of the Seiberg-Witten data onto the Calogero-Moser 
data is certainly natural, there is no direct proof of it, and it is important 
to check that the results inferred from it agree with known facts from quantum 
field theory. To establish this, as well as a series of further predictions
from the correspondence, we give four theorems (the
proofs may be found in [5] for the first three theorems,
and in [24] for the last one).

\bigskip

\noindent
{\bf Theorem 4.} {\it The spectral curve equation $\det (kI-L(z))=0$ is 
equivalent to
$$
\vartheta _1 \biggl (
{1 \over 2\omega _1}(z-m{\partial \over \partial k}) \big | \tau \biggr ) H(k)=0
\eqno (6.4)
$$
where $H(k)$ is a monic polynomial in $k$ of degree $N$, whose zeros (or 
equivalently whose coefficients) correspond to the moduli of the gauge theory. 
If $H(k)=\prod_{i=1}^N(k-k_i)$, then
$$
\lim_{q\to 0}{1\over 2\pi i}\oint_{A_i}kdz=k_i-{1\over 2}m.
$$}

Here, $\vartheta _1$ is the Jacobi $\vartheta$-function, which admits a simple 
series expansion in powers of the instanton factor $q=e^{2\pi i \tau}$, so that 
the curve equation may also be rewritten as a series expansion
$$
\sum _{n \in {\bf Z}} (-)^n q ^{\half n(n-1)} e^{nz} H(k-n \cdot m) =0
\eqno (6.5)
$$
where we have set $\omega _1 = -i\pi$ without loss of generality. The series 
expansion (6.5) is superconvergent and sparse in the sense that it receives
contributions only at integers that grow like $n^2$. 

\bigskip

\noindent
{\bf Theorem 5.} {\it The prepotential of the Seiberg-Witten theory obeys a 
renormalization group-type equation that simply relates $\F$ to the 
Calogero-Moser Hamiltonian, expressed in terms of the quantum order parameters 
$a_j$
$$
a_j = {1\over 2\pi i}\oint _{A_j} d \lambda
\qquad \qquad
{\partial \F \over \partial \tau} \bigg | _{a_j } 
= H(x,p) = \half \tr L(z)^2 +C \wp (z)
\eqno (6.6)
$$
Furthermore, in an expansion in powers of the instanton factor $q=e^{2\pi i 
\tau}$, the quantum order parameters $a_j$ may be computed by residue methods
in terms of the zeros of $H(k)$.}

\bigskip

The proof of (6.6) requires Riemann surface deformation
theory [5]. The fact that the quantum order parameters may 
be evaluated by residue methods arises from the fact that $A_j$-cycles may be 
chosen on the spectral curve $\Gamma$ in such a way that they will shrink to 
zero as $q \to 0$. As a result, contour integrals around full-fledged branch 
cuts $A_j$ reduce to contour integrals around poles at single points, which may 
be calculated by residue methods only. These methods were originally developed
in [25,26]. Knowing the quantum order parameters in  terms of the zeros $k_j$ of
$H(k)=0$ is a relation that may be inverted and  used in (6.6) to obtain a
differential relation for all order instanton  corrections. It is now only
necessary to evaluate explicitly the 
$\tau$-independent contribution to $\F$, which in field theory arises from 
perturbation theory. This may be done easily by retaining only the $n=0$ and 
$n=1$ terms in the expansion of the curve (6.5), so that 
$ z= \ln H(k) - \ln H(k-m)$. The results of the calculations to two instanton 
order may be summarized in the following theorem [5].

\bigskip

\noindent
{\bf Theorem 6.} {\it The prepotential, to 2 instanton order is given by $\F = 
\F ^{({\rm pert})} + \F ^{(1)} + \F ^{(2)}$. The perturbative contribution is 
given by
$$
\F ^{({\rm pert})} = {\tau \over 2} \sum _i a_i ^2
 - {1 \over 8 \pi i} \sum _{i,j} \biggl [ (a_i -a_j)^2 \ln (a_i - a_j)^2
 - (a_i -a_j -m)^2 \ln (a_i -a_j -m)^2 \biggr ]
 \eqno (6.7a)
$$
while all instanton corrections are expressed in terms of a single 
function
$$
S_i(a) = {\prod _{j=1} ^N \big [ (a_i-a_j)^2-m^2 \bigr ] \over 
\prod _{j\not=i} (a-a_j)^2}
\eqno (6.7b)
$$
as follows}
$$
\eqalign{
\F ^{(1)} & = {q \over 2 \pi i} \sum _i S_i(a_i) \cr
\F ^{(2)} & = {q^2 \over 8 \pi i} \biggl [
  \sum _i S_i(a_i) \partial _i ^2 S_i(a_i)
  + 4 \sum _{i\not= j} {S_i(a_i)S_j(a_j) \over (a_i -a_j)^2} - { 
S_i(a_i)S_j(a_j) \over (a_i -a_j-m)^2}  \biggr ] \cr}
  \eqno (6.7c)
$$

The perturbative corrections to the prepotential of (6.7a) indeed precisely 
agree with the predictions of asymptotic freedom. The formulas
(6.7c) for the instanton corrections 
$\F^{(1)}$ and $\F^{(2)}$ are new, as they
have not yet been computed by direct field theory methods.
Perturbative expansions of the prepotential in powers
of $m$ have also been obtained in [27].

\bigskip

The moduli $k_i$, $1\leq i\leq N$, of the gauge theory are 
evidently integrals of motion
of the system. To identify these integrals of motion, denote by $S$ be
any subset of $\{1,\cdots,N\}$, and let $S^*=\{1,\cdots,N\}\setminus S$,
$\wp(S)=\wp(x_i-x_j)$ when $S=\{i,j\}$. Let also $p_S$ denote
the subset of momenta $p_i$ with $i\in S$. We have [24]

\bigskip

\noindent
{\bf Theorem 7}. {\it For any $K$, $0\leq K\leq N$,
let $\sigma_K(k_1,\cdots,k_N)=
\sigma_K(k)$ be the $K$-th symmetric polynomial of $(k_1,\cdots,k_N)$,
defined by $H(u)=\sum_{K=0}^N(-)^K\sigma_K(k)u^{N-K}$. Then}
$$
\sigma_K(k)
=
\sigma_K(p)
+
\sum_{l=1}^{[K/2]}m^{2l}
\sum_{|S_i\cap S_j|=2\delta_{ij}\atop 1\leq i,j\leq l}
\sigma_{K-2l}(p_{(\cup_{i=1}^lS_i)^*})
\prod_{i=1}^l[\wp(S_i)+{\eta_1\over\omega_1}]
\eqno(6.8)
$$

\bigskip
 
\noindent
{\bf c) Partial Decoupling of the Hypermultiplet and Product Gauge Groups}

The spectral curves of certain gauge theories can be easily derived
from the Calogero-Moser curves by a partial decoupling of
the hypermultiplet. Indeed, 

$\bullet$ the masses of the gauge multiplet
and hypermultiplet are $|a_i-a_j|$ and $|a_i-a_j+m|$. In suitable limits,
some of these masses become $\infty$, and states with infinite mass decouple.
The remaining gauge group is a subgroup of $SU(N)$. 

$\bullet$ When the effective coupling of a gauge subgroup
is $0$, the dynamics freeze and the gauge states become non-interacting.

\medskip
Non-trivial decoupling limits arise when $\tau\to\infty$ and $m\to \infty$.
When all $a_i$ are finite, we obtain the pure Yang-Mills theory.
When some hypermultiplets masses remain finite, the $U(1)$
factors freeze, the gauge group $SU(N)$ is broken down to
$SU(N_1)\times\cdots\times SU(N_p)$,
and the remaining hypermultiplets are in e.g. fundamental or
bifundamental representations. For example,
let $N=2N_1$ be even, and set
$$
k_i=v_1+x_i,\ \
k_{N_1+j}=v_2+y_j,
\ \
1\leq i,j\leq N_1,
$$
with $\sum_{i=1}^{N_1}x_i=\sum_{j=1}^{N_1}y_j=0$. (The term $v=v_1-v_2$
is associated to the $U(1)$ factor of the gauge group).
In the limit $m\to \infty$, $q\to 0$, with $x_i$, $y_j$, $\mu=v-m$
and $\Lambda=mq^{1\over N}$ kept fixed,
the theory reduces to a $SU(N_1)\times SU(N_1)$ gauge theory,
with a hypermultiplet in the bifundamental
$(N_1,\bar N_1)\oplus (\bar N_1,N_1)$, and
spectral curve
$$
A(x)-t(-)^{N_1}B(x)
-2^{N_1}\Lambda^{N_1}({1\over t}-t^2)=0,
\eqno(6.9)
$$
where $A(x)=\prod_{i=1}^{N_1}(x-x_i)$, $B(x)=\prod_{j=1}^N(x+\mu-y_j)$,
$t=e^z$. This agrees with the curve found by Witten [28]
using M Theory, and by Katz, Mayr, and Vafa [29] using
geometric engineering.

\medskip
The prepotential of the $SU(N_1)\times SU(N_1)$ theory can be
also read off the Calogero-Moser prepotential. It is convenient to
introduce $x_i^{(I)}$, $I=1,2$, by $x_i^{(1)}=x_i$, $x_i^{(2)}=y_i$,
$1\leq i\leq N_1$. Set
$$
A_i^I=\prod_{j\not=i\atop j\in I}(x-x_j^{(I)}),
\ \
B^I(x)=\prod_{j\in J\atop |I-J|=1}(\mu\pm(x-x_j^{(J)})),
\ \
S_i^I(x)={B^I(x)\over A_i^I(x)^2},
$$
where the $\pm$ sign in $B^I(x)$ is the same
as the sign of $J-I$. Then the the first two
orders of instanton corrections to the prepotential
for the $SU(N_1)\times SU(N_1)$ theory are given by
$$
\eqalign{\F_{SU(N_1)\times SU(N_1)}^{(1)}
=&
{(-2\Lambda)^{N_1}\over 2\pi i}\sum_{I=1,2}\sum_{i\in I}S_i^I(x_i^{(I)})\cr
\F_{SU(N_1)\times SU(N_1)}^{(2)}
=&
{(-2\Lambda)^{2N_1}\over 8\pi i}\sum_{I=1,2}\sum_{i\in I}S_i^I(x_i^{(I)})
{\partial^2S_i^I(x_i^{(I)})\over\partial x_i^{(I)2}}
+
\sum_{i\not=j\atop i,j\in I}
{S_i^I(x_i^{(I)})S_j^I(x_j^{(I)})\over (x_i^{(I)}-x_j^{(I)})^2}\cr}.
\eqno(6.10)
$$
We note that an alternative derivation of (6.4) was recently presented in [30].

\bigskip
\bigbreak

\centerline{\bf VII. CALOGERO-MOSER AND SEIBERG-WITTEN THEORY}
\centerline{\bf FOR GENERAL $\G$}

\bigskip
We consider now the $\N=2$ supersymmetric gauge theory for
a general simple gauge algebra $\G$ and 
a hypermultiplet of mass $m$ in the adjoint representation. Then [9]

\medskip

$\bullet$ the Seiberg-Witten curve of the theory
is given by the spectral curve $\Gamma=\{(k,z)\in{\bf C}\times\Sigma;
\det(kI-L(z))=0\}$ of the {\it twisted} elliptic
Calogero-Moser system associated to the Lie algebra $\G$.
The Seiberg-Witten differential $d\lambda$ is given by $d\lambda=kdz$.

\medskip
$\bullet$ The function $R(k,z)=\det(kI-L(z))$ is polynomial
in $k$ and meromorphic in $z$. The spectral curve $\Gamma$
is invariant under the Weyl group of $\G$. It depends
on $n$ complex moduli, which can be thought
of as independent integrals of motion of the Calogero-Moser system.

\medskip
$\bullet$
The differential $d \lambda=kdz$ is meromorphic on $\Gamma$,
with simple poles. The position and residues of the poles are
independent of the moduli. The residues are linear
in the hypermultiplet mass $m$. (Unlike the case of
$SU(N)$, their exact values
are difficult to determine for general $\G$).

\medskip
$\bullet$ In 
the $m \to 0$ limit, the Calogero-Moser system reduces to
a free system, the spectral curve $\Gamma$ is just the producr
of several unglued copies of the base torus $\Sigma$,
indexed by the constant eigenvalues of $L(z)=p\cdot h$.
Let $k_i$, $1\leq i\leq n$, be
$n$ independent eigenvalues, and $A_i,B_i$ be the $A$
and $B$ cycles lifted to the corresponding sheets.
For each $i$, we readily obtain 
$$
\eqalign{
a_i & ={1\over 2\pi i}\oint_{A_i}d\lambda
={k_i\over 2\pi i}\oint_Adz={2\omega_1\over 2\pi i}k_i
\cr
a_{Di} & ={1\over 2\pi i}\oint_{B_i}d\lambda
={k_i\over 2\pi i}\oint_Bdz={2\omega_1\over 2\pi i}\tau k_i
\cr}
$$
Thus the prepotential $\F$ is given by
$\F={\tau\over 2}\sum_{i=1}^na_i^2$. This is the classical prepotential
and hence the correct
answer, since in the $m\to 0$ limit, the theory
acquires an $\N=4$ supersymmetry, and receives no
quantum corrections.

\medskip
$\bullet$ The $m\to\infty$ limit is the crucial consistency check,
which motivated the introduction of the {\it twisted} Calogero-Moser
systems in the first place [7,8]. In view of Theorem 2
and subsequent comments, in the limit $m\to\infty$, $q\to 0$,
with $x=X+2\omega_2{1\over h_{\G}^{\vee}}\rho$,
$m=Mq^{-{1\over 2h_{\G}^{\vee}}}$ with $X$ and $M$ kept fixed,
the Hamiltonian and spectral curve for the twisted
elliptic Calogero-Moser system with Lie algebra $\G$
reduce to the Hamiltonian and spectral curve for the
Toda system for the affine Lie algebra $(\G^{(1)})^{\vee}$.
This is the correct answer. Indeed,
in this limit, the gauge theory with adjoint hypermultiplet
reduces to the pure Yang-Mills theory,
and the Seiberg-Witten spectral curves for
pure Yang-Mills with gauge algebra $\G$ have been shown by
Martinec and Warner [31] to be
the spectral curves of the Toda system
for $(\G^{(1)})^{\vee}$.

\medskip
$\bullet$ The effective prepotential can be evaluated explicitly
in the case of $\G=D_n$ for $n\leq 5$. Its logarithmic singularity
does reproduce the logarighmic singularities expected
from field theory considerations.

\medskip
$\bullet$ As in the known correspondences between Seiberg-Witten theory
and integrable models [5,25], we expect the following equation to hold
$$
{\partial \F\over\partial\tau}=H_{\G}^{twisted}(x,p),
\eqno(6.11)
$$
to hold. Note that the left hand side can be
interpreted in the gauge theory as a renormalization group equation.

\medskip
$\bullet$ For simple laced $\G$, the curves $R(k,z)=0$ are modular
invariant. Physically, the gauge theories for these Lie algebras
are self-dual. For non simply-laced $\G$,
the modular group is broken to the congruence subgroup
$\Gamma_0(2)$ for $\G=B_n,C_n$, $F_4$, and to $\Gamma_0(3)$
for $G_2$. The Hamiltonians of the twisted Calogero-Moser systems
for non-simply laced $\G$ are also transformed under Landen
transformations into the Hamiltonians of the twisted Calogero-Moser system
for the dual algebra $\G^{\vee}$. It would be interesting to determine whether
such transformations exist for the
spectral curves or the corresponding gauge theories themselves.

\medskip

Spectral curves for certain gauge theories with classical gauge algebras and
matter in the adjoint representation have also
been proposed in [32] and [33], based on branes and M-theory. Relations with integrable systems were discussed in [34].

\bigskip
\bigbreak

\centerline{{\bf ACKNOWLEDGMENTS}}

\bigskip

We are happy to acknowledge invitations to lecture on the material presented in this paper by Harry Braden and Igor Krichever at the workshop ``Integrability, the Seiberg-Witten and Whitham Equations" at Edinburgh, Sepember 1998, by Ryu Sasaki and Takeo Inami at the ``Workshop on Gauge Theory and Integrable Models" at Kyoto, January 1999, and by Tohru Eguchi and Norisuke Sakai at ``Supersymmetry and Unified Theory of Elementary Particles" at Kyoto, February 1999. We wish to thank these organizers for their warm hospitality and for their generous support.

\bigskip
\bigbreak

\centerline{\bf REFERENCES}
\bigskip

\item{[1]} N. Seiberg and E. Witten,
``Electro-magnetic duality, monopole condensation,
and confinement in $\N=2$ supersymmetric Yang-Mills theory",
Nucl. Phys. {\bf B 426} (1994) 19-53, hep-th/9407087;
``Monopoles, duality, and chiral symmetry breaking
in $\N=2$ supersymmetric QCD",
Nucl. Phys. {\bf B 431} (1994) 494, hep-th/9410167.

\item{[2]} A. Gorskii, I.M. Krichever, A. Marshakov, A. Mironov and A. Morozov,
``Integrability and Seiberg-Witten exact solution", Phys. Lett. {\bf B355}
(1995) 466, hep-th/9505035; T. Nakatsu and K. Takasaki, ``Whitham-Toda
Hierarchy and $\N=2$ supersymmetric Yang-Mills Theory", Mod. Phys. Lett. {\bf
A11} (1996) 157, hep-th/9509162.

\item{[3]} R. Donagi and E. Witten,
``Supersymmetric Yang-Mills and integrable systems",
Nucl. Phys. {\bf B 460} (1996) 288-334, hep-th/9510101.

\item{[4]} A. Gorsky and N. Nekrasov, ``Elliptic Calogero-Moser System
from two-dimensional current algebra", hep-th/9401021; N. Nekrasov,
``Holomorphic bundles and many-body systems", Comm. Math. Phys. {\bf 180}
(1996) 587;
E. Martinec, ``Integrable structures in supersymmetric gauge
and string theory", hep-th/9510204.

\item{[5]} E. D'Hoker and D.H. Phong,
``Calogero-Moser systems in $SU(N)$ Seiberg-Witten theory",
Nucl. Phys. {\bf B 513} (1998) 405-444, hep-th/9709053.

\item{[6]} M.A. Olshanetsky and A.M. Perelomov,
``Completely integrable Hamiltonian systems
connected with semisimple Lie algebras",
Inventiones Math. {\bf 37} (1976) 93-108;
``Classical integrable finite-dimensional
systems related to Lie algebras",
Phys. Rep. {\bf 71 C} (1981) 313-400.

\item{[7]} E. D'Hoker and D.H. Phong,
``Calogero-Moser Lax pairs with spectral parameter
for general Lie algebras", Nucl. Phys. {\bf B 530} (1998)
537-610, hep-th/9804124.

\item{[8]} E. D'Hoker and D.H. Phong,
``Calogero-Moser and Toda systems for twisted and
untwisted affine Lie algebras",
Nucl. Phys. {\bf B 530} (1998) 611-640, hep-th/9804125.

\item{[9]} E. D'Hoker and D.H. Phong,
``Spectral curves for super Yang-Mills with adjoint
hypermultiplet for general Lie algebras",
Nucl. Phys. {\bf B 534} (1998) 697-719, hep-th/9804126.

\item{[10]} B. de Wit, A. Van Proeyen, Nucl. Phys. {\bf B245} (1984) 89;
S. Ferrara, Mod. Phys. Lett. {\bf A6} (1991) 2175;
A. Strominger, Comm. Math. Phys. {\bf 133} (1990) 163.

\item{[11]}   W. Lerche, 
``Introduction to Seiberg-Witten theory and its stringy origins",
Proceedings of the {\it Spring School and Workshop
on String Theory}, ICTP, Trieste (1996),
hep-th/9611190, Nucl. Phys. Proc. Suppl. {\bf B 55} (1997) 83.

\item{[12]} I.M. Krichever and D.H. Phong,
``Symplectic forms in the theory of solitons",
hep-th/9708170, to appear in Surveys in Differential Geometry, Vol. III.

\item{[13]} R. Donagi, ``Seiberg-Witten integrable models", alg-geom/9705010;
R. Carroll, ``Prepotentials and Riemann surfaces", hep-th/9802130.

\item{[14]} A. Marshakov,
``On integrable systems and supersymmetric gauge
theories", Theor. Math. Phys. {\bf 112} (1997)
791-826, hep-th/9702083.

\item{[15]} F. Calogero,
``Exactly solvable one-dimensional many-body problems",
Lett. Nuovo Cim. {\bf 13} (1975) 411-416.

\item{[16]} J. Moser, 
``Three integrable Hamiltonian systems connected with
isospectral deformations",
Advances Math. {\bf 16} (1975) 197.

\item{[17]} H. Braden, ``A conjectured R-matrix",
J. Phys. A {\bf 31} (1998) 1733-1741.

\item{[18]} A.P. Polychronakos, ``Generalized Statistics in one dimension",
hep-th/9902157.

\item{[19]} I.M. Krichever, 
``Elliptic solutions of the Kadomtsev-Petviashvili equation
and integrable systems of particles",
Funct. Anal. Appl. {\bf 14} (1980) 282-290.

\item{[20]} H.W. Braden and V.M. Buchstaber,
``The general analytic solution
of a functional equation of addition type",
Siam J. Math. anal. {\bf 28} (1997) 903-923.

\item{[21]} A. Bordner, E. Corrigan, and R. Sasaki,
``Calogero-Moser systems: a new formulation", hep-th/9805106;
A. Bordner, R. Sasaki, and K. Takasaki,
``Calogero-Moser systems II: symmetries and foldings", hep-th/9809068;
A. Bordner and R. Sasaki, ``Calogero-Moser
systems III: Elliptic potentials and twisting", hep-th/9812232.

\item{[22]} I. Inozemtsev, 
``Lax representation with spectral parameter on a torus for integrable
particle systems", Lett. Math. Phys. {\bf 17} (1989) 11-17;
``The finite Toda lattices", Comm. Math. Phys. {\bf 121} (1989) 628-638.

\item{[23]} P. Goddard and D. Olive,
``Kac-Moody and Virasoro algebras in relation to
quantum physics", International J. Mod. Phys. A, Vol. I (1986)
303-414.

\item{[24]} E. D'Hoker and D.H. Phong,
``Order parameters, free fermions, and
conservation laws for Calogero-Moser systems",
hep-th/9808156, to appear in Asian J. Math.

\item{[25]} E. D'Hoker, I.M. Krichever, and D.H. Phong,
``The renormalization group equation for $\N=2$
supersymmetric gauge theories",
Nucl. Phys. {\bf B 494} (1997) 89-104, hep-th/9610156.

\item{[26]} E. D'Hoker I.M. Krichever and D.H. Phong, 
``The effective prepotential for $\N=2$ supersymmetric $SU(N_c)$ gauge
theories", Nucl. Phys. {\bf B489} (1997) 179, hep-th/9609041.

\item{[27]} J. Minahan, D. Nemeschansky, and N. Warner,
``Instanton expansions for mass deformed $\N=4$ super Yang-Mills
theory", hep-th/9710146.

\item{[28]} E. Witten, 
``Solutions of four-dimensional field theories via
M-Theory", Nucl. Phys. {\bf B 500} (1997) 3, hep-th/9703166.

\item{[29]} S. Katz, P. Mayr, and C. Vafa,
``Mirror symmetry and exact solutions of 4D
$\N=2$ gauge theories", Adv. Theor. Math. Phys.
{\bf 1} (1998) 53, hep-th/9706110.

\item{[30]} K. Vaninsky,
``On explicit parametrization of spectral curves
for Moser-Calogero particles and its applications",
December 1998 preprint.

\item{[31]} E. Martinec and N. Warner,
``Integrable systems and supersymmetric gauge theories",
Nucl. Phys. {\bf B 459} (1996) 97-112, hep-th/9509161.

\item{[32]} A.M. Uranga, 
``Towards mass deformed $\N=4$ $SO(N)$ and $Sp(K)$ gauge
theories from brane configurations",
Nucl. Phys. {\bf B 526} (1998) 241-277, hep-th/9803054.

\item{[33]} T. Yokono, 
``Orientifold four plane in brane configurations and $\N=4$ $USp(2N)$
and $SO(2N)$ theory",
Nucl. Phys. {\bf B 532} (1998) 210-226, hep-th/9803123.

\item{[34]} A. Gorsky, ``Branes and Integrability in the $\N=2$ SUSY
YM theory", Int. J. Mod. Phys. {\bf A12} (1997) 1243, hep-th/9612238;
A. Gorsky, S. Gukov, A. Mironov, ``SUSY field theories, integrable 
systems and their stringy brane origin", hep-th/9710239;
A. Cherkis and A. Kapustin, ``Singular monopoles and supersymmetric
gauge theories in three dimensions", hep-th/9711145.

\end